\documentclass{nature}
\usepackage{graphicx}
\usepackage[caption = false]{subfig}
\usepackage{academicons}
\usepackage{amsmath, amssymb}
\usepackage{graphicx}
\usepackage{graphicx}
\usepackage{hyperref}
\usepackage{threeparttable}
\usepackage{xcolor}
\usepackage{ulem}
\usepackage{rotating}

\usepackage{caption}
\newcommand{\insight}{\textit{Insight}-HXMT}

\title{Insight-HXMT identification of a non-thermal X-ray burst from SGR J1935+2154 and with FRB 200428}

\author{
C.K. Li$^{1}$\thanks{Co-First Authors. These authors contributed equally and are in alphabetical order:
Cheng-Kui Li, Lin Lin and Shao-Lin Xiong}, L. Lin$^{2*}$, S.L. Xiong$^{1*}$, M.Y. Ge$^{1}$, X.B. Li$^{1}$, T.P. Li$^{1,3,4\dag}$, F.J. Lu$^{1\dag}$, S.N. Zhang$^{1,3}$\thanks{Co-Corresponding Authors. These authors contributed equally and are in alphabetical order:
Ti-Pei Li, Fang-Jun Lu and Shuang-Nan Zhang},  Y.L. Tuo$^{1,3}$, Y. Nang$^{1,3}$, B. Zhang$^{5}$
% ...
}

\begin{document}

\maketitle
\begin{affiliations}
\item Key Laboratory of Particle Astrophysics, Institute of High Energy Physics, Chinese Academy of Sciences, 19B Yuquan Road, Beijing 100049, China
\item Department of Astronomy, Beijing Normal University, Beijing 100875, China
\item University of Chinese Academy of Sciences, Chinese Academy of Sciences, Beijing 100049, China
\item Department of Astronomy, Tsinghua University, Beijing 100084, China
\item Department of Physics and Astronomy, University of Nevada, Las Vegas, NV 89154, USA
\item Technology and Engineering Center for Space Utilization, Chinese Academy of Sciences, Beijing 100094, China
\item Institut f\"{u}r Astronomie und Astrophysik, Kepler Center for Astro and Particle Physics, Eberhard Karls Universit\"{a}t, Sand 1, 72076 T\"{u}bingen, Germany
\item Institute of High Energy Physics, Chinese Academy of Sciences, 19B Yuquan Road, Beijing 100049, China
\item Department of Physics, Tsinghua University, Beijing 100084, China
\item Department of Engineering Physics, Tsinghua University, Beijing 100084, China
\item Computing Division, Institute of High Energy Physics, Chinese Academy of Sciences, 19B Yuquan Road, Beijing 100049, China
\item Key Laboratory of Space Astronomy and Technology, National Astronomical Observatories, Chinese Academy of Sciences, Beijing 100012, China
\item College of physics Sciences and Technology, Hebei University, Baoding City, Hebei Province, 071002, China
\item School of Physics and Optoelectronics, Xiangtan University, Xiangtan City, Hunan Province, 411105, China
\item College of Physics, Jilin University, Changchun City, Jilin Province, 130012, China
\end{affiliations}

\begin{abstract} 

Fast radio bursts (FRBs) are short pulses observed in radio band from cosmological distances\cite{lorimer07}. One class of models invoke soft gamma-ray repeaters (SGRs), or magnetars, as the sources of FRBs\cite{petroff19}. Some radio pulses have been observed from some magnetars\cite{camilo06}, however, no FRB-like events had been detected in association any magnetar burst, including one giant flare\cite{tendulkar16}. Recently, a pair of FRB-like bursts (FRB~200428 hereafter) separated by 29 milliseconds (ms) were detected from the general direction of the Galactic magnetar SGR~J1935+2154\cite{2020Natur.587...54T,2020Natur.587...59B}. Here we report the detection of a non-thermal X-ray burst in the 1--250\,keV energy band with the \textit{Insight}-HXMT satellite\cite{zhangsn2020}, which we identify as emitted from SGR~J1935+2154. The burst showed two hard peaks with a separation of 34 ms, broadly consistent with that of the two bursts in FRB~200428. The delay time between the double radio and X-ray peaks is $\sim8.62$ s, fully consistent with the dispersion delay of FRB~200428. We thus identify the non-thermal X-ray burst is associated with FRB~200428 whose high energy counterpart is the two hard peaks in X-ray. Our results suggest that the non-thermal X-ray burst and FRB~200428 share the same physical origin in an explosive event from SGR~J1935+2154.

\end{abstract}

SGR~J1935+2154 was discovered when it went into outburst in 2014\cite{israel2016}. Since then and before 2020, the source experienced several activities in 2015 February, 2016 May to July and 2019 November\cite{2017ApJ...847...85Y,lin2020,2020ApJ...902L..43L}. Between outbursts, isolated bright flares or short bursts in X-ray or gamma-ray have been detected from the source\cite{kozlova2016,lin2020}. These make SGR~J1935+2154 one of the most active magnetars. Starting from 2020 April 27 18:26:20 UT, a series of X-ray and gamma-ray instruments were triggered by multiple short bursts and a burst forest including hundreds of bursts from SGR~J1935+2154\cite{SGR-Swift,2020arXiv200907886Y}. Within thirteen hours, we started a long Target of Opportunity (ToO) observation of this source using \textit{Insight}-HXMT with all its three collimated telescopes covering 1--10\,keV (Low Energy X-ray telescope, LE), 5--30\,keV (Medium Energy X-ray telescope, ME) and 20--250\,keV (High Energy X-ray telescope, HE), respectively. This pointed ToO observation continued for 60~ks from April 28 07:14:52 UT to April 29 11:53:01 UT.  
During the \textit{Insight}-HXMT observation, a double-peaked and short radio burst, FRB~200428, from the general direction of SGR~J1935+2154 was reported by CHIME/FRB\cite{2020Natur.587...54T} and STARE2\cite{2020Natur.587...59B} at April 28 UTC 14:34:24 (at 600 MHz and 1.4\,GHz), respectively. The fluence of this radio burst recorded by STARE2\cite{2020Natur.587...59B} is $\sim$1.5~MJy~ms, which is over six magnitudes brighter than those radio bursts from XTE~J1810-197\cite{camilo06}, which had been the brightest radio bursts from magnetars. This makes it the first possible magnetar radio burst detectable from an extra-galactic distance (e.g FRB 180916.J0158+65 at 149\,Mpc\cite{2020Natur.587...59B}), if FRB~200428 were emitted from SGR~J1935+2154.

\begin{figure*}
\begin{center}
\caption{Localization of the burst using \textit{Insight}-HXMT HE, ME and LE data. The red crosses mark the known position of SGR~J1935+2154.  The white contours in the zoomed-in panel are $1\sigma$, $2\sigma$ and $3\sigma$ uncertainty regions of the localization with \textit{Insight}-HXMT data. The best position of this burst is $\sim$3.7 arcmin away from SGR~J1935+2154 with $1\sigma$ error of $\sim$10 arcmin (see {\bf Methods} for details about localization). The red circle and blue-dotted ellipse mark the 95\% localization regions of FRB~200428 determined by CHIME/FRB\cite{2020Natur.587...54T} and STARE2\cite{2020Natur.587...59B}, respectively.}
\includegraphics[width=0.75\textwidth,clip]{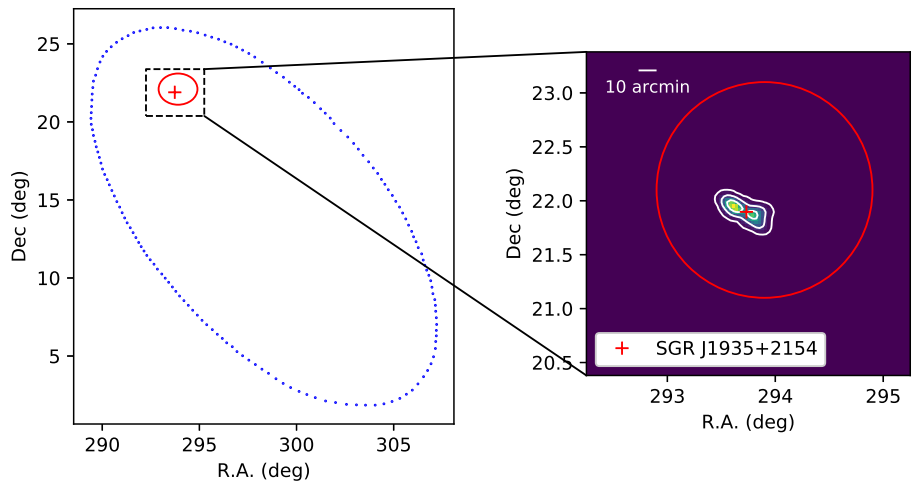}
\label{fig:location}
\end{center}
\end{figure*}

\textit{Insight}-HXMT detected a series of 11 bursts within about 17 hours of exposure to SGR~J1935+2154 (see {\bf Methods} for description and Supplementary Table 1 for burst list). It is mostly likely that most, if not all, of these bursts came from SGR~J1935+2154, since it was the only active magnetar in this period and in the field of view of \textit{Insight}-HXMT. The brightest burst with a trigger time (denoted as $T_0$) of April 28 14:34:24.0000 UT (satellite time) or 14:34:24.0114 UT (geocentric time) lasted for about 1 second in 1--250~keV and was seen clearly in all three telescopes. This burst is also the closest one in time to FRB~200428. With different orientations of the collimators, \textit{Insight}-HXMT can localize the burst within its field of view, as shown in Figure \ref{fig:location}. The burst is located at ${\rm RA=293.67^{+0.16}_{-0.11}}$\,deg, ${\rm Dec=21.92^{+0.08}_{-0.07}}$\,deg, $\sim$3.7 arcmin away from SGR~J1935+2154 with $1\sigma$ error of $\sim$10 arcmin. We thus identify this burst as coming from SGR~J1935+2154. INTEGRAL was also triggered on this event and identified its source to the same magnetar\cite{2020ApJ...898L..29M}.

\begin{figure*}
\begin{center}
\caption{The lightcurves and the spectrum evolution during the burst of SGR~J1935+2145 observed with \textit{Insight}-HXMT. The reference time is $T_0$ (2020-04-28 14:34:24 UTC). The vertical dashed lines indicate two peaks in the lightcurves. The separation between the two lines are ~30\,ms. \textbf{(a)}: The lightcurve observed with \textit{Insight}-HXMT/HE with a time resolution of 1\,ms near the peak and 10\,ms outside the peak. Due to the saturation effect, there are bins near the peak with no photons recorded for both HE and LE. \textbf{(b)} and \textbf{(c)} are the lightcurves observed with ME and LE with a time bin of 5\,ms, respectively. The bottom three panels show the spectral evolution of the burst. \textbf{(d)}, \textbf{(e)} and \textbf{(f)} are photo index $\gamma$, $E_{\rm {cut}}$ and $E_{\rm {peak}}$ of the CPL model. The errors are given at the 68\% confidence level.(see \textbf{Methods} for details of the saturation and the deadtime correction.)}
\includegraphics[width=0.8\textwidth]{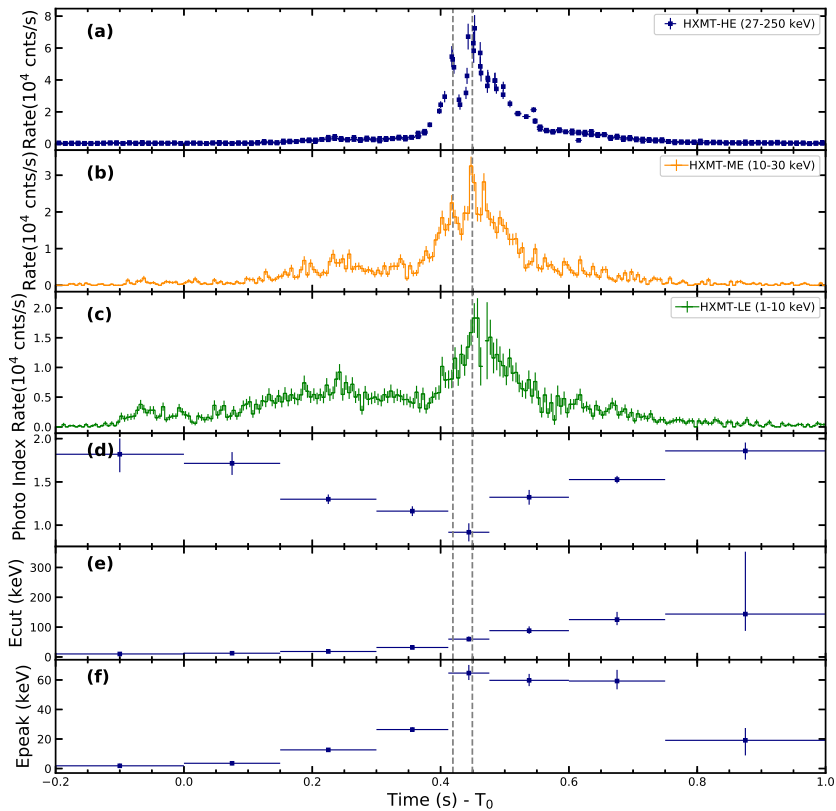}
\label{fig:lightcurve}
\end{center}
\end{figure*}

This burst was so bright that it saturated both LE and HE, and also caused moderate deadtime effects in ME. After correcting all these effects (see {\bf Methods}), the lightcurves of the burst obtained by the three telescopes are presented in Figure \ref{fig:lightcurve}. The full lightcurves of this burst consist of two major bumps separated by about 0.2~s, and a minor soft bump just before $T_0$ that is only present in LE and ME data, indicating overall spectral evolution as shown  in Figure \ref{fig:lightcurve} during the burst. The second major bump, which was also detected by INTEGRAL\cite{2020ApJ...898L..29M} and Konus-Wind\cite{2020arXiv200511178R}, is much brighter than the first one. In the lightcurves of both ME and HE, two narrow peaks are clearly seen (see {\bf Methods}) during the second major bump. In the LE lightcurve, only the second narrow peak is visible significantly, indicating somewhat different broad band energy spectra between the two narrow peaks. The separation time between the two narrow X-ray peaks ($\sim34$~ms) is broadly consistent with the 29 ms separation between the two narrow peaks in FRB~200428, and the apparent time lag between X-ray and radio peaks ($\sim$8.62~s) is in good agreement with the calculated dispersion delay (8.62~s) between X-ray and radio using the DM ($\sim$333 pc/cm$^3$) measured by CHIME/FRB\cite{2020Natur.587...54T} and STARE2\cite{2020Natur.587...59B}. We thus identify the burst detected by \textit{Insight}-HXMT is associated with FRB~200428 and both belong to a single explosive event from SGR~J1935+2154.

The time-integrated spectrum of this burst ($T_0-0.2$~s to $T_0+1.0$~s) is derived jointly using HE, ME and LE data (Figure 3, see {\bf Methods} for details of spectral fitting). The best fit and statistically acceptable model is a cutoff power-law (CPL) with neutral hydrogen column density $n_{\rm H}=(2.79_{-0.17}^{+0.18})\times10^{22}~{\rm cm}^{-2}$, photon index $\Gamma=1.56\pm0.06$ and cutoff energy $E_{\rm cut}=83.89_{-7.55}^{+9.08}~{\rm keV}$ (corresponding to a peak energy $E_{\rm peak}=(2-\Gamma)E_{\rm cut}\sim37$~keV).
The unabsorbed fluence is ($7.14_{-0.38}^{+0.41})\times10^{-7}~{\rm erg}~{\rm cm}^{-2}$ in 1--250~keV, corresponding to a total emission energy of $\sim1\times 10^{40}~{\rm erg}$ for the 12.5~kpc\cite{kothes2018} distance of SGR~J1935+2154 and $\sim1\times 10^{39}~{\rm erg}$ for a closer distance of 4.4~kpc\cite{2020ApJ...898L..29M}. This burst is brighter than $\sim84\%$ of events collected from the source during $2014-2016$ with {\it Fermi}/GBM\cite{lin2020}. We also fit the spectrum with several other spectral models, e.g., single power-law (PL), double blackbody (BB+BB) and blackbody plus power-law (BB+PL). The fit to the BB+PL model is marginally consistent with data, with slightly higher column density ($n_{\rm H}=(3.50\pm0.17)\times10^{22}~{\rm cm}^{-2}$) and larger photon index ($\Gamma=1.93\pm0.04$); the flux of the unabsorbed blackbody component with temperature of $11.32_{-0.56}^{+0.55}~{\rm keV}$ is only 18\% of the total flux in 1--250\,keV. The other two models provide significantly worse fit and are thus rejected. 

We conclude that the integrated spectrum is dominated by a power-law covering at least the 1-100 keV range, and thus this burst is primarily non-thermal in nature. It is also clear that the two narrow peaks separated by 34~ms must also be dominated by a non-thermal spectrum, since the peak energy ($E_{\rm {peak}}$ in time resolved spectral analysis) reaches its maximum during the peak of the second bump of the lightcurves where the two narrow peaks are found. It is interesting to note that the lower limit of the radio flux detected with STARE2\cite{2020Natur.587...59B} falls in between the extrapolated values from the non-thermal X-ray spectrum with the power-law parameters of the fits to the CPL and BB+PL models (see the inset (f) in Figure 3).

In summary, with the observation of \textit{Insight}-HXMT we have identified that the short non-thermal X-ray burst was emitted by the Galactic magnetar SGR~J1935+2154 and produced almost simultaneously with FRB~200428 in a single explosive event. In the literature, FRB emission has been interpreted as either coherent curvature radiation of electron-positron pairs from a neutron star magnetosphere\cite{katz14,kumar17,yangzhang18} or synchrotron maser emission in a relativistic, magnetized shock\cite{lyubarsky14,plotnikov19}. Since magnetar bursts are believed to be magnetosphere-related\cite{thompson95}, the fact that the narrow double peaks in both radio and X-ray are emitted around the same time, and hence, likely originate from the same emission region, lends support to the magnetospheric models of FRBs.

However, a thermal origin is preferred for normal short bursts from magnetars\cite{israel2008,lin2012}. We notice that $\sim6\%$ of the bursts ($7/109$) from SGR~J1935+2154 detected with Fermi/GBM between 2014 and 2016 can be best fit with a power-law model\cite{lin2020}. The fluence of these bursts is about one order of magnitude dimmer than this one associated with FRB~200428. We therefore set a conservative upper limit of $6\%$ to the percentage of magnetar bursts which may have similar radio emission to FRB~200428. This is consistent with the non-radio detection from the hard X-ray bursts of SGR~J1935+2154 during the same bursting phase\cite{2020Natur.587...63L}. Actually, non-thermal X-ray bursts are very rarely observed from magnetars in general, which may explain why events similar to FRB~200428 have not been seen previously. However, selection bias cannot be excluded since large field of view radio observations have only become available recently.

Previously we have conducted a search for prompt $\gamma$-ray counterparts to FRBs\cite{2020A&A...637A..69G} in the \textit{Insight}-HXMT data and obtained only upper limits as low as $5.5\times 10^{47}$ erg s$^{-1}$ over 1 s for the periodic repeater FRB 180916.J0158+65. If this X-ray burst were emitted from an extragalactic magnetar located at a distance of FRB~180916.J0158+65 at 149 Mpc\cite{marcote20}, and assume the distance of SGR~J1935+2154 is 12.5~kpc\cite{kothes2018}, then the observed fluence should be $\sim4\times10^{-15}~{\rm erg}~{\rm cm}^{-2}$ in 1-250 keV ($\sim5\times10^{-16}~{\rm erg}~{\rm cm}^{-2}$ for a distance of 4.4~kpc\cite{2020ApJ...898L..29M}), which is far below the sensitivity limits of the X-ray telescopes currently in orbit (or in the foreseeable future). This may explain the non-detection of the X-ray counterpart of any cosmological FRB so far. Nevertheless, our identification of FRB~200428 with a magnetar means at least some of FRBs are produced by magnetars, thus FRBs can be used as an effective tool to study the extra-galactic magnetars, which are otherwise undetectable.  On the other hand, giant flares from magnetars can have peak luminosity of $10^{44-47}~{\rm erg}~{\rm s}^{-1}$, about 4--7 orders of magnitude more luminous than this non-thermal X-ray burst, and thus might be detectable with the current X-ray telescopes in orbit or the future X-ray missions, such as eXTP\cite{extp} which has a much larger effective area in the X-ray band than those X-ray telescopes in orbit.

\clearpage

\begin{figure*}
\begin{center}
\caption{The spectrum observed with \textit{Insight}-HXMT covers the 1--250\,keV energy band. Data from the three telescopes of \textit{Insight}-HXMT covering different energy bands are represented in different colors (LE: black, ME: red and HE: green). In the fitting process, we introduced a constant factor to offset the different saturation and deadtime effects in different detectors. Four models were considered, cutoff power-law (CPL), blackbody+power-law (BB+PL), power-law (PL), and blackbody+blackbody (BB+BB). The equivalent hydrogen column in the interstellar absorption model was free to fit. \textbf{(a)} The X-ray spectrum of SGR~J1935+2154 described by CPL model. The inset (\textbf{f}) shows the comparison between the radio flux lower limit detected with STARE2\cite{2020Natur.587...59B} and extrapolations from the X-ray spectrum to the radio frequency range, where the green and orange regions are the $3\sigma$ error bands with the parameters of the CPL (below STARE2) and BB+PL (above STARE2) models, respectively. Panels \textbf{(b)}-\textbf{(e)} are the residuals of the data from the individual models, respectively. The errors are given at the 68\% confidence level.(see {\bf Methods} for details of the spectral fitting and parameters derived.)}
\includegraphics[width=0.6\textwidth]{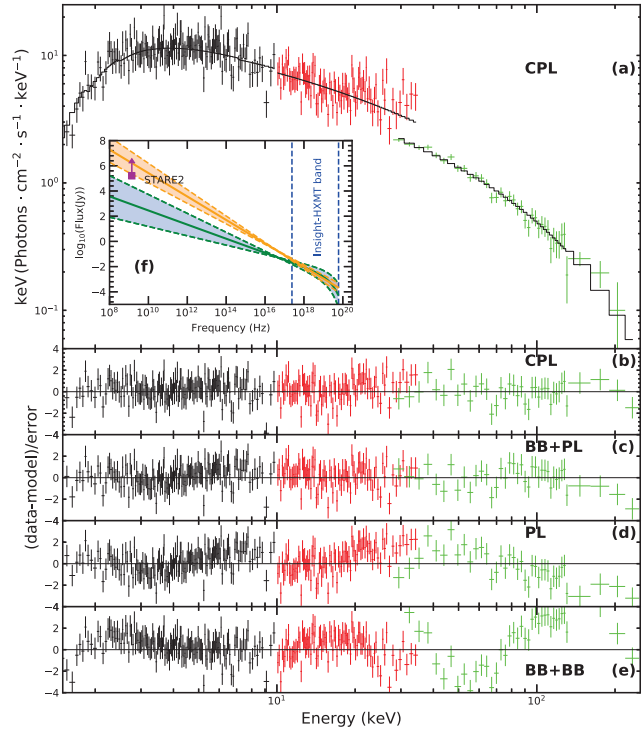}
\label{fig:spectrum}
\end{center}
\end{figure*}

\clearpage

\section*{Methods}

\subsection{\textit{Insight}-HXMT observations and burst search}

The \textit{Insight}-Hard X-ray Modulation Telescope (\textit{Insight}-HXMT) is China's first X-ray astronomy satellite\cite{zhangsn2020, S.Zhang..2018, li2020flight} which was launched on June 15th, 2017. It has an altitude of 550\,km and an inclination of 43\,degrees. As a broadband X-ray (1--250 keV) observatory, \textit{Insight}-HXMT consists of three telescopes, i.e., the High Energy X-ray telescope (HE) using 18 NaI(Tl)/CsI(Na) phoswich scintillation detectors for 20--250 keV\cite{Lcz2019}, the Medium Energy X-ray telescope (ME) using 1728 Si-PIN detectors for 5--30\,keV\cite{Cxl2019}, and the Low Energy X-ray telescope (LE) using 96 Swept Charge Device (SCD) detectors for 1--15\,keV\cite{Cy2019}. All three telescopes use slat collimators to confine their Field Of Views (FOVs). In addition to the pointed or scanning observation with the collimators, \textit{Insight}-HXMT can also monitor the all-sky in gamma-ray (0.2--3 MeV) using the CsI scintillation detectors of HE. More details about the \textit{Insight}-HXMT can be found in\cite{zhangsn2020}.

A dedicated and long Time of Opportunity (ToO) observation of \textit{Insight}-HXMT was implemented for SGR~J1935+2154 from 2020-04-28T07:14:51 to 2020-04-29T00:00:00, and a throughout search for X-ray bursts have been made.  The trigger condition for the search is that the count rates of three or more NaI detectors of HE exceeds the background count rate, which is the mean count rate in the previous 10 s, with significance greater than $3\sigma$ at five time scales ( 0.05\,s, 0.1\,s, 0.2\,s, 0.5\,s and 1\,s). This search results in 11 bursts. The starting time and other properties are listed in Supplementary Table 1, where the fluence is obtained by fitting their spectra with simple spectral models (i.e. PL or CPL), as the fluence does not varies significantly with which spectral model is used. Saturation and deadtime corrections are made before spectral fitting, according to the procedures described below. More detailed analyses of these bursts will be presented elsewhere. 

The rest of this {\bf Methods} part is mainly dedicated to the burst at 2020-04-28T14:34:24.00 (UTC) that is associated with FRB~200428. Because of the extreme brightness, the \textit{Insight}-HXMT data suffers substantial saturation and deadtime effects, which require dedicated corrections as detailed below.

\subsection{Data analysis}

The timing and spectral results of the X-ray burst associated with FRB~200428 are obtained by analysing the \textit{Insight}-HXMT  1L data with the \textit{Insight}-HXMT Data Analysis Software package (\texttt{HXMTDAS}) version 2.02. Specifically, the steps are: (1) Use the commands \texttt{hepical}, \texttt{mepical}, \texttt{lepical} in \texttt{HXMTDAS} to calibrate the photon events from the 1L data according to the Calibration Database (CALDB) of \textit{Insight}-HXMT. As for HE, the short spikes with known characteristics produced in the electronics are removed from the physical events. (2) Select the good time intervals (GTIs) from $T_0$-0 to $T_0$+1\,s, where $T_0$ is 2020-04-28 14:34:24 UTC. (3) Extract the good events based on the GTIs using the commands \texttt{hescreen}, \texttt{mescreen}, and \texttt{lescreen}. (4) Generate the spectrum with the selected events using the commands \texttt{hespecgen}, \texttt{mespecgen}, and \texttt{lespecgen}.  (5) Create the background spectrum from the events in the time interval ${T}_0-51$ to ${T}_0-1\,\rm{s}$. (6) Generate the response matrix files required for spectral analysis using the commands \texttt{herspgen}, \texttt{merspgen}, and \texttt{lerspgen}. (7) Produce the raw ME and LE lightcurves using the commands, \texttt{melcgen}, and \texttt{lelcgen}. 

Due to the strong saturation effect in both LE and HE data, the raw data in some time intervals were discarded on-board and their lightcurves need to be corrected as presented below.

\subsection{Data saturation and deadtime correction}

Because of the extremely high flux of the burst, the detected events exceeded the storage limits of their on-board data buffer, and so the observed data suffered from saturation. The observational effect of saturation is that in some time intervals the events are lost.  Besides the saturation effect, during the procession of an event by the front-end electronics, the detectors sharing the same Physical Data Acquisition Unit (PDAU) can not record any photons, and such an effect is called deadtime. As will be detailed below, both HE and LE suffered strongly from the saturation effects, while the deadtime effects are significant in the HE and ME data. Both the saturation and deadtime need to be corrected when we produce the lightcurves and spectra.

The 18 phoswich X-ray detectors of HE are divided into three groups, each contains six detectors that share one PDAU.
Therefore, the three groups of detectors have different event-lost intervals and the common gaps for three groups are shown in Supplementary Figure 1. We correct for the saturation effects in the data of the three groups independently, and then combine them together when we produce the final lightcurve.

The steps of saturation correction for a group of HE detectors are listed as follows:
(1) Find the start and stop time of the intervals in which the raw data are not lost.
(2) Calculate the deadtime ratio of each detector as a function of time, the details of which can 
be found in Xiao et al. (2020)\cite{Xiao2020}. 
(3) Screen the data in these time intervals to discard the CsI events (anti-coincident events), as well as the events whose energies are out of the selected energy band. Then, the number of NaI events can be obtained for each detector in the group. Using the time intervals selected in the first step and the deadtime ratio calculated in the second step, the true source count rate of each detector in the group can be obtained. 
(4) Merge the count rate of all detectors in each group and calculate the error of the merged rate. It should be noted that, for the third group (Group ID is 2), as the events of the blinded detector are not used, a factor of 6/5 is used to normalize its count rate, so that the count rates of the three groups can be compared at the same level and combined together to produce the overall HE lightcurve.

According to the design of HE, the detected events are sent to a First-In-First-Out (FIFO) storage buffer. Limited by the FIFO memory capacity and the data transmission bandwidth on board, part of data blocks would be lost when observing ultra bright sources. That is why we can find some gaps in the original light curves. Fortunately, thanks to the special status flags of the FIFO, we can recover some data points with average count rate values or absolute upper/lower limit values to fill the gaps in the raw count rates, consisting of events of all energies between 20-250 keV but without information on the energy deposition of each event. As shown in Supplementary Figure 2, the structure of the two narrow peaks is visible more clearly in the HE raw count rates during the X-ray burst. From the upper limits of the raw count rates corresponding to the first peak of the FRB, the data gaps do not affect either the position of first narrow peak or the separation of the double narrow peaks of the X-ray burst.

ME does not suffer from the saturation effect and the raw data have no time gap. The deadtime of ME can be calculated with HXMTDAS v2.02. The number and ratio of the lost events in $T_{0}+0.37$ and $T_{0}+0.62\, \mathrm{s}$ are also listed in Supplementary Table 2.
The lightcurves before and after deadtime corrections are shown in Supplementary Figure 3. The lightcurve in energy range 18-50 keV is shown in Supplementary Figure 2. There are three narrow peaks in this light cure. The first two narrow peaks correspond to the two narrow peaks of HE lightcurve. While the third narrow peak around T0+0.47s, corresponding to the second bright peak in whole lightcurve in Figure \ref{fig:lightcurve}(b) (10-30 keV), is softer, which means that most of the X-ray photons have lower energy and caused the data loss of LE and the gaps in the LE lightcurve as shown in Figure \ref{fig:lightcurve}(c) or Supplementary Figure 4.

LE has three detector boxes and each box contains 32 SCD detectors. The data of each detector box 
can be processed independently. 
In the LE data, besides the normal physical events with energies
above the on-board threshold, LE also has the forced trigger
events, which record the amplitude of the noise or the pedestal
offset for each SCD detector in every 32\,ms\cite{li2020flight}. 
The count rate of the forced trigger events in each detector box is 1000 counts per second if there is no saturation effect. 

The LE lightcurves are then corrected for saturation using the count rate of the recorded forced trigger events. Since the three detector boxes have different saturated time intervals, we reconstructed the LE lightcurve with almost the full time coverage. The lightcurves before and after saturation correction are shown in Supplementary Figure 4.
The deadtime of LE caused by the force trigger events can also be calculated by HXMTDAS, which are listed in Supplementary Table 2. It is a minor and negligible issue.

\subsection{The two narrow peaks}

As shown in Figure \ref{fig:lightcurve}, the lightcurve in each energy band roughly consists of two bumps located at around $T_0$+0.2 and $T_0$+0.45\,s, and the HE and ME lightcurves show two narrow peaks on the second main bump. In order to estimate the significance and to get the exact time of each peak, the HE and ME lightcurves are fitted by five Gaussian functions, in which two of them are used to describe the two narrow peaks, 
\begin{equation}
    R=N_{\rm p1}G(t,t_{\rm p1},\sigma_{\rm p1}) + N_{\rm p2}G(t,t_{\rm p2},\sigma_{\rm p2})+R_{\rm 3},
    \label{eq1}
\end{equation}
where $G(t,t_{\rm p},\sigma_{\rm p})=\frac{1}{\sqrt{2\pi}\sigma_{\rm p}}\exp(-\frac{(t-t_{\rm p})^2}{2\sigma_{\rm p}^{2}})$, $N_{\rm p1}$ and $N_{\rm p2}$ are the normalization, $t_{\rm p1}$ and $t_{\rm p2}$ are the arrival times of the two narrow peaks, $\sigma_{\rm p1}$ and $\sigma_{\rm p2}$ are the Gaussian widths of the two narrow peaks. $R_{\rm 3}=\sum_{i=3}^{5}{G(t,t_{\rm pi},\sigma_{\rm pi})}+l$ describes the three Gaussian functions for the broad components of the lightcurve, where $l$ is the background level of the lightcurve. From the fitting results, the separation $\tau$ of the two narrow peaks is calculated from $t_{\rm p2}-t_{\rm p1}$.

As shown in Supplementary Figure 5 (a) and (b), the lightcurves of HE and ME could be well fitted by equation \ref{eq1}. If the normalization of the two narrow components is set to 0, the reduced-$\chi^{2}$ is 4.1 (d.o.f.=30) for the fitting to the data points in $T_0$+0.35 to $T_0$+0.43\,s that contains the first narrow peak. Similarly, the reduced-$\chi^{2}$ is 3.0 (d.o.f.=29) for duration $T_0$+0.43 to $T_0$+0.50\,s that contains the second narrow peak. These large reduced-$\chi^2$ values verify the high detection significance of the two narrow peaks. 

As shown in Supplementary Figure 5 (c), the lightcurve of LE can be well fitted by $R=N_{\rm p2}G(t,t_{\rm p2},\sigma_{\rm p2})+R_{\rm 3}$. A narrow peak corresponding to the second narrow peak in HE and ME lightcurves is also visible, though not as significant as in HE and ME lightcurves.

\subsection{Spectral analyses and model comparison}
We extract the spectrum using data in a duration of 1.2 s, from $T_0$ - 0.2\,s to $T_0$ + 1\,s. Deadtime correction is a built-in function of the \texttt{HXMTDAS} and has been considered in spectral analysis. However, the saturation correction is not implemented in spectrum generation but will be dealt with in spectral fitting process.

We use XSPEC version 12.10.0c to analyze the spectra. Four different models are used to fit the spectra, which are
(1) single power-law (PL), (2) cutoff power-law (CPL) , (3) two blackbody (BB+BB) and (4) blackbody plus power-law (BB+PL). In addition, we use a constant ($const$) to represent the saturation effect in LE and HE and the $wabs$ model to account for the absorption of the interstellar medium.  Eventually, the four models are:
$wabs*cutoffpl*const$, $wabs*pow*const$, $wabs*(bb+bb)*const$ and $wabs*(bb+pow)*const$. The best-fit parameters and their uncertainties are listed
in Supplementary Table 4. The distribution of the fitted residuals is displayed in Figure 2 of the main article. 

From Supplementary Table 4 and Figure 2 of the main article we can easily reject the 
single power-law model and the two temperature blackbody model, but the cutoff power-law (CPL) and the blackbody plus power-law model (BB+PL)
fit the burst spectra well, though the latter has relatively higher $\chi^2$ values and slightly structured residual above
80 keV. Discussions about these models can be found in the main article.

Supplementary Table 5 lists the spectrum parameters of the burst compared with Konus-Wind and INTEGRAL. When using all data of LE, ME and \&HE, the reduced $\chi^2$ indicates that the CPL model is much better than the 2BB model, while Konus-Wind and INTEGRAL data can not distinguish the two models.
For comparison, we limit the energy band of the spectrum as from 20\,keV to 250\,keV, and re-extract the spectrum in time range from ~$T_0$+0.2~s to ~$T_0$+0.8~s (similar to INTEGRAL) or from ~$T_0$+0.435~s to ~$T_0$+0.7~s (similar to Konus-Wind). In such cases, we obtain similar spectrum parameters as Konus-Wind and INTEGRAL, and can not determine which model is preferred. The unabsorbed flux is ($8.08_{-0.43}^{+0.45})\times10^{-7}~{\rm erg}~{\rm s}^{-1}~{\rm cm}^{-2}$ in 20--200~keV with the CPL model, which is slightly lower than the one obtained by INTEGRAL ($10.05_{-0.5}^{+0.5}\times10^{-7}~{\rm erg}~{\rm s}^{-1}~{\rm cm}^{-2}$), possibly due to cross-calibration issues between the two instruments. As mentioned in the main article, we can exclude the thermal spectral model (BB+BB) and conclude the non-thermal nature of this peculiar X-ray burst, due to the broad energy coverage of \textit{Insight}-HXMT, especially the 1-20 keV coverage which is not available in either Konus-Wind or INTEGRAL.

\subsection{Time resolved spectrum}

In order to analyze the spectral evolution, we split the GTI from $T_0-0.2$~s to $T_0+1$~s into eight time intervals and extract the spectrum from each of them (Supplementary Table 6). We fit the eight spectra with cut-off power-law model and $pgstat$ statistics. The neutral hydrogen column density is fixed to the same value of the time-integrated spectrum. The parameter $factor$ of the component $const$ in the model is set free due to the data saturation and deadtime effects.
The evolution of $E_{\rm {peak}}$, photon index $\Gamma$ and $E_{\rm cut}$ for the burst associated with the FRB are shown in the bottom three panels of Figure \ref{fig:lightcurve}.
$E_{\rm cut}$ increases with time, $\Gamma$ reaches the minimum value at the two peaks, while $E_{\rm peak}$ arises before the two peaks, keeps a plateau during the burst, and falls down to a low value after the burst.

\subsection{Localization of the X-ray burst}

Although the three telescopes of \textit{Insight}-HXMT point to the same nominal directions, the long axis directions of their Field Of Views (FOVs) are different, which could be used to locate the burst. Supplementary Figure 6 shows the FOVs of the three telescopes of \textit{Insight}-HXMT. Every telescope has three groups of FOVs whose long axis directions are 60 degree different from the neighbouring ones. When the direction of a source deviates from the center of the FOVs, the count rates on detectors with different FOVs decrease with different slopes, following the shapes of the Point Spread Functions (PSF) \cite{Nang2020}, which allow us to fit the position of the source using the count rates of the burst on different detectors and their PSFs and thus provides localization of bright sources within the FOV of \textit{Insight}-HXMT with several arcmin accuracy \cite{Nang2020}.

PSFs of all \textit{Insight}-HXMT collimators have been calibrated \cite{Nang2020}, which are then used to reconstruct the position of the source from the differences of the count rates between different FOVs. This localization method has been extensively tested and verified with \textit{Insight}-HXMT observations \cite{SAI20201,GUAN202011}.

For the localization of this burst, count rates of all the three telescopes from UTC 2020-04-28T14:34:24 to UTC 2020-04-28T14:34:25 are used, after saturation and deadtime corrections according to Supplementary Table 2. In the fitting, for all the the three telescopes the same burst position (RA and Dec) parameters are assumed with three different normalized flux parameters. A Markov Chain Monte Carlo (MCMC) algorithm is utilized in the fitting. The best fitting result gives a reduced $\chi^2$ of 0.845 for 4 degrees of freedom. Figure \ref{fig:location} shows the distributions of position parameters derived from the MCMC approach. The best-fit location of the burst is 3.7 arcmin away from that of SGR~J1935+2154 with $1\sigma$ uncertainty of 10 arcmin, fully consistent with SGR~J1935+2154.

{\bf Data availability}

The data that support the plots within this paper and other findings of this study are available from the \textit{Insight}-HXMT website (http://www.hxmt.cn/ or http://www.hxmt.org/).

{\bf Code Availability}

The \textit{Insight}-HXMT data reduction was performed using software available from the \textit{Insight}-HXMT website (http://www.hxmt.cn/ or http://www.hxmt.org/). The model fitting of spectra was completed with XSPEC, which is available from the HEASARC website (https://heasarc.gsfc.nasa.gov/).

\begin{addendum}

 \item[Acknowledgements] This work made use of the data from the \textit{Insight}-HXMT mission, a project funded by China National Space Administration (CNSA) and the Chinese Academy of Sciences (CAS). The \textit{Insight}-HXMT team gratefully acknowledges the support from the National Program on Key Research and Development Project (Grant No. 2016YFA0400800) from the Minister of Science and Technology of China (MOST) and the Strategic Priority Research Program of the Chinese Academy of Sciences (Grant No. XDB23040400). The authors thank supports from the National Natural Science Foundation of China under Grants U1838105, U1838111, U1838113, U1838202, 11473027, 11733009, U1838201, 1173309, U1838115, U1938109, Y829113, 11673023, U1838104, 11703002

\item[Author Contributions]  C.-K.L., L.L. and S.-L.X. are co-first authors and listed in alphabetical order. T.-P.L., F.-J.L. and S.-N.Z. are co-corresponding authors and listed in alphabetical order. T.-P.L. was the initial proposer and PI of \textit{Insight}-HXMT. S.-N.Z. is the current PI of \textit{Insight}-HXMT since 2016, organized the observations, data analysis and presentation of the results, writing and editing of the paper. L.L. proposed the ToO observation, is a main writer of the paper and participated in discussions. S.-L.X. participated in organizing the observations, data analysis, discussion and paper writing. F.-J.L. is a leader in building \textit{Insight}-HXMT and participated in organizing the data analysis, discussions and paper writing. C.-K.L. is the main contributor to the data analysis and participated in paper writing. B.Z. is responsible for theoretical interpretation, and participated in organizing the observations, discussions,  and paper writing. M.-Y.G., Y.-L.T., X.-B.L., Y.N., S.X., Y.C., L.-M.S., Y.T., X.-F.Z., C.-Z.L., S.-M.J., J.-Y.L. and B.L. participated in the data analysis and discussion. All other authors contributed to developing, building and operating the \textit{Insight}-HXMT payloads and science data center.

\item[Competing Interests] The authors declare no competing interests.

\item[Correspondence] Correspondence and requests for materials should be addressed to (T.-P.L., F.-J.L. and S.-N.Z., email:  litp@ihep.ac.cn, lufj@ihep.ac.cn, zhangsn@ihep.ac.cn).

\end{addendum}

\clearpage
\setcounter{figure}{0}
\setcounter{table}{0}

\captionsetup[figure]{labelfont={bf},labelformat={default},labelsep=period,name={Supplementary Figure}}

\captionsetup[table]{labelfont={bf},labelformat={default},labelsep=period,name={Supplementary Table}}

\begin{table}
\footnotesize
\caption {Bursts detected by \insight{} from 2020-04-28T07:14:51 to 2020-04-29T00:00:00. In the table, trigger time is the satellite time, the energy band for fluence calculation is 1--250\,keV, duration is that covers 90\% of the burst counts, and $\Delta t$ is the time difference between burst and FRB 200428.  The errors are given at the 68\% confidence level.}
\scriptsize{}
\label{tab:burst_list}
\medskip
\begin{center}
\begin{tabular}{c  r c r }
\hline \hline
Trigger time (UTC)  & Fluence & Duration &$\Delta t$ \\
&${\rm 10^{-8}erg~cm^{-2}}$&s&s\\
\hline
2020-04-28T08:03:34.35	&${\rm	5.65 	\pm	1.14 	}$&	0.11 	&	-23458.65 	\\
2020-04-28T08:05:50.15	&${\rm	5.04 	\pm	1.39 	}$&	0.07 	&	-23322.85 	\\
2020-04-28T09:08:44.30	&${\rm	1.37 	\pm	1.86 	}$&	0.06 	&	-19548.70 	\\
2020-04-28T09:51:04.90	&${\rm	25.58 	\pm	2.51 	}$&	0.42 	&	-17008.10 	\\
2020-04-28T11:12:58.55	&${\rm	1.30 	\pm	1.41 	}$&	0.06 	&	-12094.45 	\\
2020-04-28T12:54:02.20	&${\rm	0.87 	\pm	1.09 	}$&	0.40 	&	-6030.80 	\\
2020-04-28T14:20:52.50	&${\rm	2.93 	\pm	1.17 	}$&	0.60 	&	-820.50 	\\
2020-04-28T14:20:57.90	&${\rm	2.06 	\pm	2.45 	}$&	0.06 	&	-815.10 	\\
2020-04-28T14:34:24.00	&${\rm	63.68 	\pm	6.62 	}$&	0.53 	&	-9.00 	\\
2020-04-28T17:15:26.25	&${\rm	0.25 	\pm	0.42 	}$&	0.08 	&	9653.25 	\\
2020-04-28T19:01:59.85	&${\rm	3.01 	\pm	1.22 	}$&	0.16 	&	16046.85 	\\

\hline \hline
\end{tabular}
\end{center}
\end{table}

\begin{table}
\footnotesize
\caption{Events lost due to saturation and deadtime in $T_{0}+0.37$  and $T_{0}+0.62\, \mathrm{s}$ }
\scriptsize{}
\label{table:datalossratio}
\medskip
\begin{center}
\begin{tabular}{c c c c c c}
\hline \hline
Telescope  & Group ID &  N1$^{\rm a}$ & LR1$^{\rm b}$ & N2$^{\rm c}$ & LR2$^{\rm d}$  \\

\hline
    &  0 & 5627 & 66.0\% & 981  &11.5\% \\
 HE &  1 & 6210 & 70.8\% &  1106 & 12.6\%\\
    &  2 & 4793 &  61.7\% & 909  & 11.7\%\\
\hline
     &  0  &  0   &  0  &  379 & 32.8\% \\
 ME  &  1  &  0   &  0  & 554  & 47.6\% \\
     &  2  &  0   &  0  &  688 & 53.0\% \\
\hline
    &  0 &  276  & 29.6\%   &  0.26 &  0.03\%  \\
 LE & 1  &  377  &  35.2\%  &  0.27 &  0.03\% \\
    & 2  &  418  & 37.6\%   &  0.27 &  0.03\% \\
\hline \hline
\end{tabular}
\\
$^{\rm a}$  N1 is the number of events lost due to saturation. \\
$^{\rm b}$  LR1 is the lost ratio of events due to saturation. \\
$^{\rm c}$  N2 is the number of events lost due to deadtime. For LE, the deadtime is induced by the forced trigger events. \\
$^{\rm d}$  LR2 is the lost ratio of events due to deadtime.
\end{center}
\end{table}


\begin{figure}[htbp]
    \centering
    \includegraphics[width=0.6\textwidth]{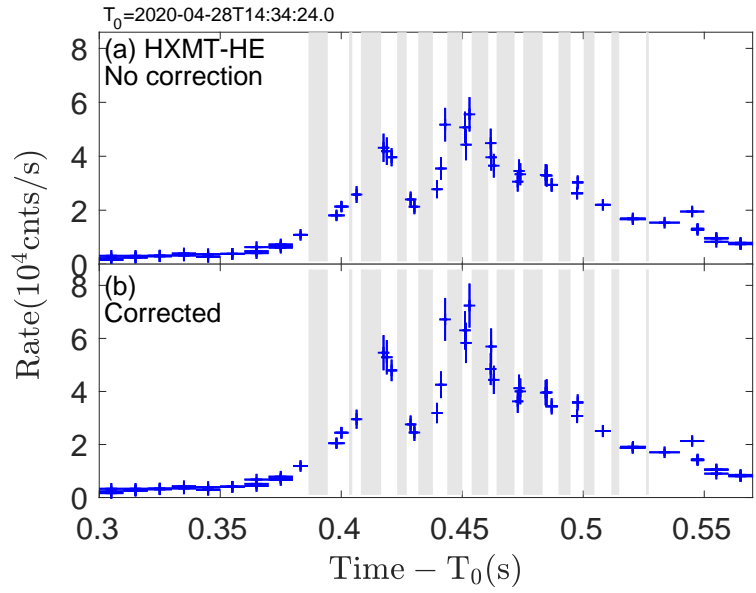}
    \caption{The lightcurves of HE. Panel (a): lightcurve before deadtime correction.  Panel (b): lightcurve after deadtime correction. The gray belts represent time intervals for data loss. The errors are given at the 68\% confidence level.}
    \label{fig:helcbox0}
\end{figure}

\begin{figure}[htbp]
    \centering
    \includegraphics[width=0.6\textwidth]{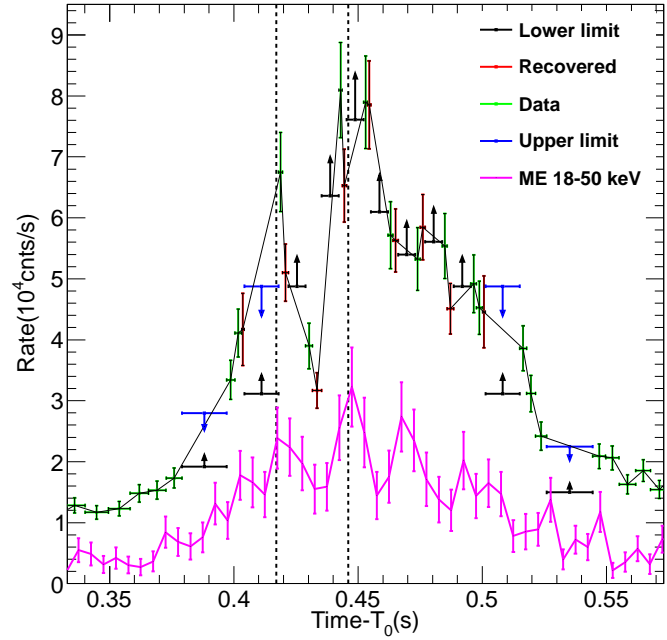}
    \caption{The recovered lightcurve of HE group 0 from raw data. The black and blue points are recovered absolute lower limits and upper limits from raw data, respectively. The red points are recovered data from the raw data. The light curve of ME between 18 and 50\,keV are also displayed but the rate has been re-scaled arbitrarily in order to be compared with HE. The errors are given at the 68\% confidence level.}
    \label{fig:helcbox0_rec}
\end{figure}

\begin{figure}[htbp]
    \centering
    \includegraphics[width=0.5\textwidth]{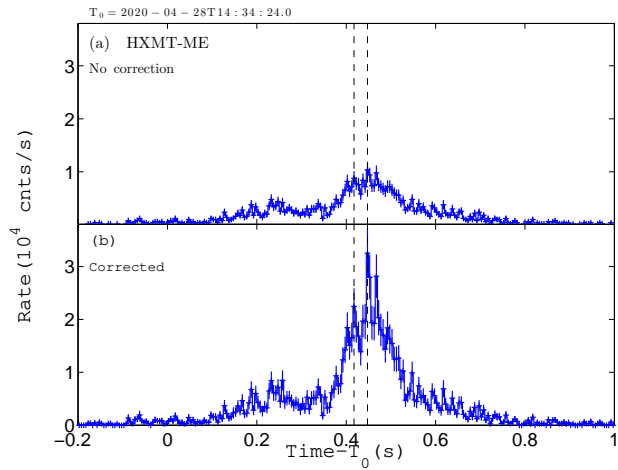}
    \caption{The lightcurves of ME. Panel (a): lightcurve without deadtime correction. Panel (b): lightcurve after deadtime correction. The errors are given at the 68\% confidence level.}
    \label{fig:melc}
\end{figure}

\begin{figure}[htbp]
    \centering
    \includegraphics[width=0.5\textwidth]{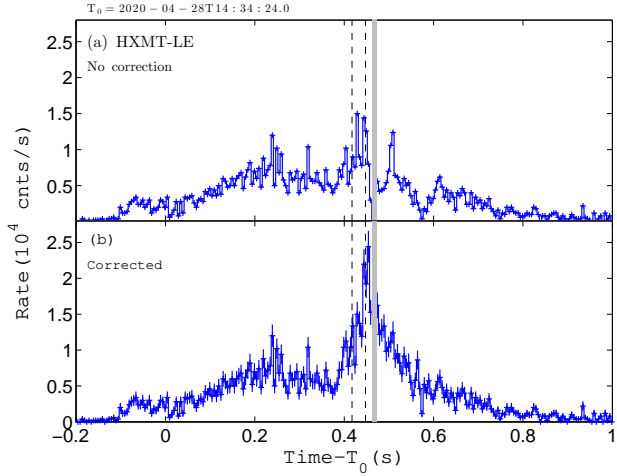}
    \caption{The lightcurves of LE. Panel (a): lightcurve before correction of lost events. Panel (b): light curve after lost events correction. The gray belts represent the time interval in which none of the three detector boxes was recording photon events normally. The errors are given at the 68\% confidence level.}
    \label{fig:lelc}
\end{figure}

\begin{figure}[htbp]
    \centering
    \includegraphics[width=0.5\textwidth]{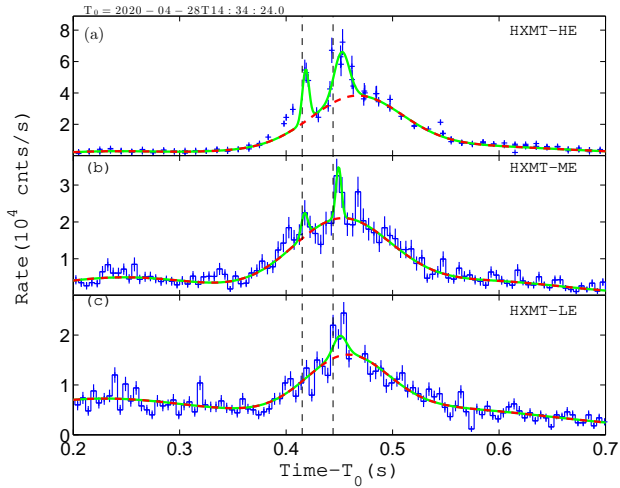}
    \caption{Fitting to the lightcurves. The blue points are lightcurves obtained from \insight{} HE/ME/LE. The vertical dashed lines are the arrival times of the narrow peaks. The red lines represent the sum of the three broad Gaussian functions. In panels (a) and (b), the green lines represent the fitted curves with the sums of the five Gaussian functions for ME and HE, in which two are for the two narrow peaks. In panel (c), the green line represents the fitted curve to the LE lightcurve with four Gaussian functions, in which one is for the narrow peak in coincidence to the second peak in HE and ME lightcurves. The errors are given at the 68\% confidence level.}
    \label{fig:lcfit}
\end{figure}

\begin{table}
\footnotesize
\caption{Fitting parameters of the two narrow peaks for HE and ME, and one narrow (second) peak for LE.  The errors are given at the 68\% confidence level.}
\scriptsize{}
\label{table:clfit}
\medskip
\begin{center}
\begin{tabular}{c c c c c c}
\hline \hline
Telescope  & ${\rm t_{p1}}$ (ms) & $\sigma_{\rm p1}$ (ms)  & ${\rm t_{p2}}$ (ms) & $\sigma_{\rm p2}$ (ms) & ${\rm \tau}$ (ms) \\
\hline
HE   &  $418\pm2$& $3.1\pm2.7$      & $452\pm1$   &   $7.0\pm0.8$ & $34\pm2$  \\
ME   &  $419\pm2$& $3.0\pm1.7$      & $449\pm2$   &   $3\pm3$ & $32\pm3$ \\
LE   & -- & -- & $450\pm2$   &   $6\pm3$ & -- \\
\hline
\end{tabular}
\end{center}
\end{table}

\begin{sidewaystable}

\footnotesize
\caption{Best-fit free parameters of the burst. The integration time for spectrum is from $T_0$-0.2\,s to $T_0$+1\,s. Four models are employed to fit the spectrum observed by \insight{}, as cutoff power-law (CPL), power-law (PL), two blackbody (BB+BB), and a model combine blackbody and power-law (BB+PL). $n_{\rm H}$ is the equivalent hydrogen column in the model for interstellar absorption.  The errors are given at the 68\% confidence level.}
\scriptsize{}
\label{tab:spec_parameters}
\medskip
\begin{center}
\begin{tabular}{c c r c r r c c c c c c c c}
\hline \hline
Model     &  $n_{\rm H}$&    $kT_{1}$ & $kT_{2}$&$\rm{Norm}_{1}$ & $\rm{Norm}_{2}$&$\rm{PhoIndex}$  & $E_{\rm{cut}}$&  $\rm{factor}_{\rm{ME}}$&$ \rm{factor}_{\rm{HE}}$&$flux_{1}$&$flux_{2}$&$\chi^{2}/d.o.f$  \\
 & (${\rm 10^{22}cm^{-2}}$) &  (keV) & (keV) &$ $ &$ $           &  & (keV)  &  &  & ${\rm 10^{-7}erg}~{\rm cm^{-2} s^{-1}}$ & ${\rm 10^{-7}erg}~{\rm cm^{-2} s^{-1}}$ & \\
\hline
CPL	&	$2.79 _{-0.17}^{+0.18 }$	&	$--$	&	$--$	&	$31.48_{-3.13}^{+3.50 }$	&	$--$	&	$1.56 _{-0.06}^{+0.06 }$	&	$83.89_{-7.55}^{+9.08 }$	&	$0.98 _{-0.06}^{+0.07 }$	&	$0.68 _{-0.07}^{+0.07 }$ & $5.95_{-0.32}^{+0.34}$ &$--$	&	1.00/242	\\\\
PL	&	$4.26 _{-0.18}^{+0.19 }$	&	$--$	&	$--$	&	$87.26_{-4.95}^{+5.17 }$	&	$--$	&	$2.21 _{-0.03}^{+0.03 }$	&	$--$	&	$1.68 _{-0.08}^{+0.08 }$	&	$1.60 _{-0.13}^{+0.13 }$&$4.61_{-0.24}^{+0.26}$&$--$	&	1.48/243	\\\\
BB+BB	&	$0.55 _{-0.11}^{+0.12 }$	&	$1.63 _{-0.04}^{+0.04 }$	&	$14.46_{-0.24}^{+0.25 }$	&	$1.77 _{-0.04}^{+0.05 }$	&	$4.37 _{-0.42}^{+0.46 }$	&	$--$	&	$--$	&	$1.84 _{-0.16}^{+0.17 }$	&	$0.45 _{-0.04}^{+0.05 }$&$1.47_{-0.04}^{+0.04}$&$3.65_{-0.35}^{+0.39}$	&	2.14/241	\\\\
BB+PL	&	$3.50 _{-0.17}^{+0.17 }$	&	$11.32_{-0.56}^{+0.55 }$	&	$--$	&	$1.56 _{-0.27}^{+0.31 }$	&	$54.46_{-3.87}^{+4.17 }$	&	$1.93 _{-0.04}^{+0.04 }$	&	$--$	&	$1.05 _{-0.07}^{+0.08 }$	&	$0.54 _{-0.06}^{+0.07 }$&$1.31_{-0.22}^{+0.26}$&$5.80_{-0.29}^{+0.32}$&	1.05/241	\\\\
\hline \hline
\end{tabular}
\end{center}
\end{sidewaystable}

\begin{table}
\footnotesize
\caption{Time-integred spectrum compared with KW and INTEGRAL. In order to compare with KW and INTEGRAL results, the main spectrum is extracted from 0.2~s to 0.8~s (similar with INTEGRAL), and the on-burst spectrum is extracted from 0.435~s to 0.7~s (similar with KW). We also fit the spectra above 20~keV to cover the similar energy band with KW and INTEGRAL.  The errors are given at the 68\% confidence level.}
\scriptsize{}
\label{tab:speccompare}
\medskip
\begin{center}
\begin{tabular}{lccc|ccc}
\hline \hline
Model  && CPL &&& BB+BB &  \\
\hline
Instrument & index  & Epeak  &$\chi^{2}/d.o.f$ & $kT_1$ & $kT_2$ & $\chi^{2}/d.o.f$ \\
\hline

Integral/IBIS (main)&$0.7_{-0.2}^{+0.4}$&$65\pm{5}$&1.8/13&$11\pm{1.3}$&$30\pm{4}$&1.6/12\\
LE \& ME\& HE (main)&$1.34\pm{0.05}$&$48_{-1.79}^{+1.82}$&1.13/211&$1.62\pm{0.05}$&$13.4_{-0.18}^{+0.19}$&3.47/210\\
ME\& HE (20keV, main)&$0.97_{-0.12}^{+0.12}$&$51.42_{-1.83}^{+1.72}$&1.08/66&$6.12_{-0.44}^{+0.48}$&$21.89_{-1.00}^{+1.12}$&1.26/65\\
KW (on-burst)&$0.72_{-0.47}^{+0.46}$&$85_{-10}^{+15}$&49.1/46&$11_{-4}^{+3}$&$31_{-7}^{+12}$&45.6/47\\
LE\&ME\&HE (on-burst)&$1.32\pm{0.07}$&$65.01_{-3.36}^{+3.59}$&1.34/118&$1.64\pm{0.08}$&$15.82_{-0.29}^{+0.30}$&3.14/117\\
ME\&HE (20keV, on-burst)&$0.76_{-0.18}^{+0.17}$&$66.04_{-2.23}^{+2.32}$&$1.05/23$&$9.94_{-1.48}^{+1.59}$&$27.21_{-2.05}^{+2.44}$&0.91/22\\
\hline \hline
\end{tabular}
\\
The errors of KW are given at the 90\% confidence level. PG-stat statistic is applied by KW instead of $\chi^2$ statistic.
\end{center}
\end{table}

\begin{table}
\footnotesize
\caption{Spectral fit parameters for the time resolved spectra. The reference time is $T_0$ (2020-04-28 14:34:24 UTC).  The errors are given at the 68\% confidence level.}
\scriptsize{}
\label{tab:trspec2}
\medskip
\begin{center}
\begin{tabular}{ccccccc}
\hline \hline
time interval &  index & Ecut & Epeak &  $pgstat/d.o.f.$  \\
$T-T_0 (\rm{s})$ &   & (keV)  &(keV) &  \\
\hline

-0.200 - 0.000 &$ 1.82_{-0.21 }^{  +0.18} $&$10.12  _{ -2.32  }^{ +2.86    } $&$1.84   _{ -1.40   }^{+1.49  } $&  37.62   /25\\
0.000  - 0.150 &$ 1.71_{-0.13 }^{  +0.13} $&$12.52  _{ -1.69  }^{ +2.17    } $&$3.56   _{ -1.31   }^{+1.07  } $&  111.63  /79\\
0.150  - 0.300 &$ 1.30_{-0.06 }^{  +0.05} $&$18.32  _{ -1.06  }^{ +1.17    } $&$12.63  _{ -0.45   }^{+0.44  } $&  105.46  /88\\
0.300  - 0.412 &$ 1.16_{-0.06 }^{  +0.06} $&$31.77  _{ -2.19  }^{ +2.48    } $&$26.40  _{ -1.39   }^{+1.47  } $&  89.17   /86\\
0.412  - 0.476 &$ 0.92_{-0.11 }^{  +0.10} $&$59.74  _{ -7.09  }^{ +9.05    } $&$64.64  _{ -4.70   }^{+5.52  } $&  107.07  /86\\
0.476  - 0.600 &$ 1.32_{-0.09 }^{  +0.08} $&$88.18  _{ -10.95 }^{ +14.34   } $&$59.72  _{ -3.91   }^{+4.37  } $&  75.79   /87\\
0.600  - 0.750 &$ 1.53_{-0.04 }^{  +0.04} $&$125.18 _{ -18.59 }^{ +25.92   } $&$59.25  _{ -5.60   }^{+7.56  } $&  97.67   /89\\
0.750  - 1.000 &$ 1.86_{-0.10 }^{  +0.10} $&$143.78 _{ -56.37 }^{ +209.08  } $&$19.10  _{ -10.22  }^{+8.35  } $&  43.07   /50\\
\hline \hline
\end{tabular}
\\

\end{center}
\end{table}

\begin{figure}[htbp]
    \centering
    \includegraphics[width=0.6\textwidth]{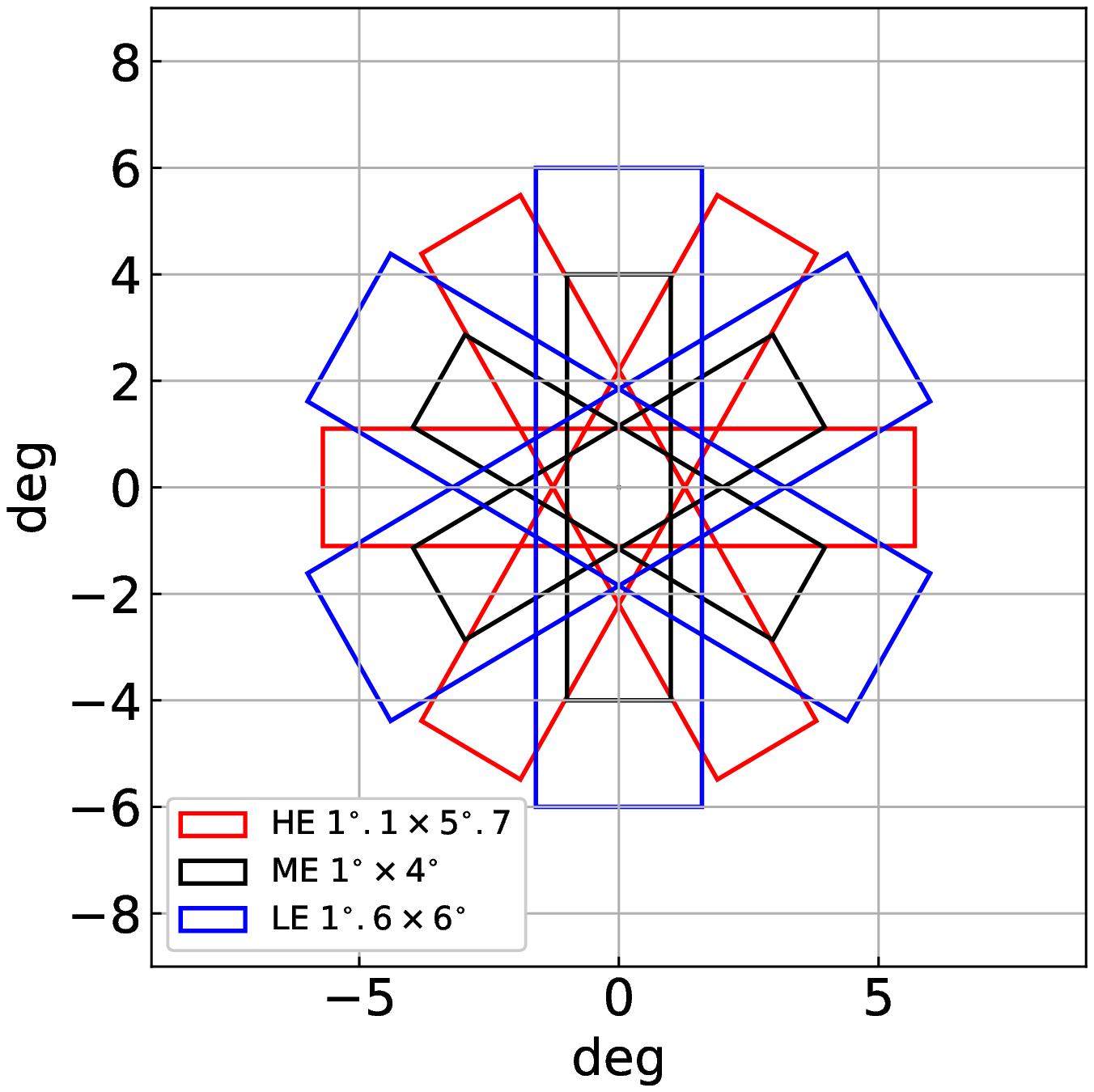}
    \caption{The FOVs of LE, ME and HE of \insight{}.}
    \label{fig:FOV}
\end{figure}

\end{document}